\documentclass[useAMS,usenatbib]{mn2e}
\usepackage{graphicx}

\def\deg{\ifmmode^\circ\else$^\circ$\fi}
\def\kms{km\thinspace s$^{-1}$}
\def\msun{M$_{\odot}$}

\title[  Embedded  Star Clusters  in  the W51  Giant  Molecular Cloud.  ]{
Embedded Star Clusters in the W51 Giant Molecular Cloud.}

\author[Kumar  M.    S.   N., Kamath U. S., and Davis, C. J. ]{M.    S.   
Nanda      Kumar$^{1}$\thanks{E-mail:nanda@astro.up.pt         (MSNK);
cdavis@jach.hawaii.edu (CJD);  kamath@crest.ernet.in (USK)}U. S. Kamath
$^{2}$\footnotemark[1] and C.  J.  Davis $^{3}$\footnotemark[1]
\\  $^{1}$Centro  de Astrofisica  da  Universidade  do  Porto, Rua das
Estrelas,  4150-762 s/n Porto,  Portugal  \\$^{2}$Indian Institute  of
Astrophysics, CREST  Campus, Shidlaghatta Road, Hosakote 562114, India
\\    $^{3}$Joint Astronomy Center,    660 N.   A'oh\={o}k\={u} Place,
University Park, Hilo, HI 96720, USA}

\begin{document}

\date{ }

\pagerange{\pageref{firstpage}--\pageref{lastpage}} \pubyear{2002}

\maketitle

\label{firstpage}

\begin{abstract}

We present sub-arcsecond (0.35\arcsec-0.9\arcsec), near-infrared J,H,K
band photometric observations of   six  fields  along the  W51   Giant
Molecular Cloud (W51 GMC).  Our observations reveal four new, embedded
clusters  and provide a  new high-resolution (0.35\arcsec) view of the
W51IRS2 (G49.5-0.4) region.  The  cluster associated with G48.9-0.3 is
found  to be  a double  cluster enclosed  in  a  nest of near-infrared
nebulosity.  We construct stellar surface density  maps for four major
clusters  in the W51   GMC.  These unveil the underlying  hierarchical
structure.  Color-color and color-magnitude diagrams for each of these
clusters show clear differences  in  the embedded stellar  populations
and indicate the relative ages  of these clusters. In particular,  the
clusters associated with the HII  regions G48.9-0.3 and G49.0-0.3  are
found to have a high fraction of YSOs and are therefore considered the
youngest  of  all the   near-infrared clusters in   the  W51 GMC.  The
estimated masses  of the  individual  clusters, when  summed, yield  a
total stellar  mass of $\sim$10$^4$\msun\ in the  W51 GMC,  implying a
star   formation efficiency of 5-10\%.    These results in comparision
with the CO observations of  the W51 GMC,  suggest for the first time,
that star formation in  the W51 GMC  is likely triggered by a galactic
spiral density wave.

\end{abstract}


\section{Introduction}

W51 is  classically known to be a  complex  of compact radio continuum
sources representing a luminous  star forming  region in the  galactic
disk.  The  $\lambda$21\,cm maps   show  two  major chunks  of   radio
emission (representing complex HII  regions)  that are referred to  as
W51A and W51B.   A  diffuse and  extended component  of radio emission
surrounding  and extending  to the  east of W51B   is  called W51C and
identified as a supernova remnant.   The HII region complexes W51A and
W51B  are associated with   intense molecular emission that is  spread
over an area of  1\deg$\times$1\deg centered  on (l,b)$\sim$(49.5\deg,
-0.2\deg). This large scale molecular emission, which encompasses both
W51A  and  W51B,  is called   the W51   Giant  Molecular Cloud  (GMC).
\citet{cs98}(hereafter  CS98)  present CO J=1--0 and  $^{13}$CO J=1--0
maps   of the W51GMC  and  show that all  of   the bright, compact HII
regions are  situated along a high  velocity cloud which they identify
as the ``68\kms'' cloud.  \citet{koo99} present higher resolution maps
of the W51B region  in CO J=2--1  and CO  J=1-0 emission.  CS98  argue
that W51 is in the top 1\%  of all Galactic  molecular clouds by size,
and is in the top 5-10\% by mass.

To better  describe  this complex region  in the  context  of the work
presented here,  we show in   Fig.1  a multi-wavelength view  of   the
W51GMC.  The top  panel  shows contours of   the $\lambda$21\,cm radio
continuum image from  \citet{km97}  overlaid on a  grey  scale 8$\mu$m
emission  image  obtained by the   MSX  mission.  The  radio continuum
sources  (filled  triangular  symbols)  G49.4-0.3  and  G49.5-0.4  are
referred  to as W51\,A and the  remaining components namely G49.2-0.3,
G49.1-0.4 and G48.9-0.3 constitute  the W51B region.  Mehringer (1994)
used   high    resolution   radio  continuum  and   recombination line
observations to  identify  several   components  of the    HII regions
associated   with  the  W51    complex.   Following  his nomenclature,
G49.5-0.4, the most luminous source in W51, is  resolved into at least
eight components  (w51a--h).  The most prominent  among these are W51d
and W51e which correspond to the strong infrared sources IRS2 and IRS1
\citep{gw94} respectively.

Fig.\,1b (bottom panel) displays $^{13}$CO J=1--0 emission contours of
the 68\kms  cloud (CS98) overlaid on a  2MASS K-band image mosaiced by
us.  Comparison of Fig.\,1a and 1b shows that while the HII regions of
W51A are fairly well enclosed  by the observed $^{13}$CO emission, the
W51B  HII regions extend beyond  the densest regions  of the molecular
cloud, traced in 8$\mu$m emission.  The mid-infrared MSX A-band image,
which  traces emission from the  large-scale  warm dust, resembles the
2MASS K-band image, which    traces the embedded stellar  content  and
hottest regions of  gas and dust, in  two ways.  The extended, whitish
regions in  the 2MASS image   (see also atlas image  of  W51 on  2MASS
website), which  are caused by  extinction  of  the background  light,
coincide with  the brighter regions (darker shades)  in the MSX image.
However,  these regions also coincide  with {\em  compact} clusters of
stars  in  the 2MASS  image.   Further, these   condensations are also
traced   by the radio   continuum emission,  which delineates the  HII
regions and therefore the ionised gas.  As is evident from Fig.1b, all
of these bright condensations,  which represent luminous  star forming
regions, are situated along the   68\kms cloud.  This indicates   that
they may be embedded  in this giant molecular  cloud.  A cursory study
of the 2MASS data showed that the regions enclosed by the bold, square
boxes  in Fig.1  likely host  embedded  clusters.   This prompted  the
observations  discussed in this  paper, where we  present high spatial
resolution near-infrared  (NIR)  photometric  studies of  the  regions
marked with  the boxes  in Fig.\,1. However,  this  is not  a complete
sample, because there are few other unstudied objects appearing on the
2MASS image that look quite similar to the regions of our choice.  The
following results provide a new  picture of large scale star formation
in the W51GMC, which was  not clear from  previous studies that mostly
concentrated on the W51A region.

\begin{figure*}

\vbox to230mm{\vfil
\resizebox{!}{8.0in}{\includegraphics[]{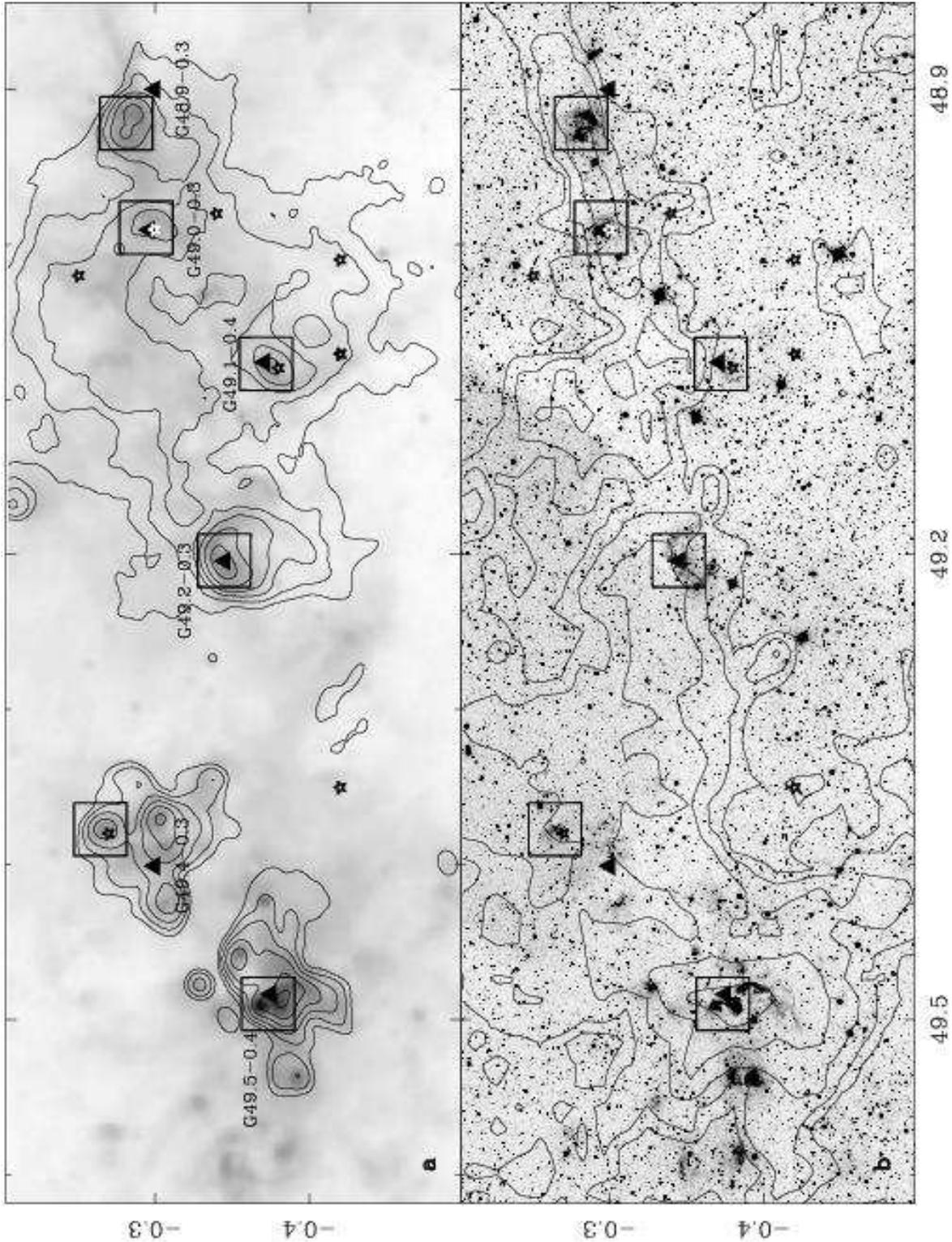}}
 \caption{a) MSX A-band (8$\mu$m) image of the W51~GMC, displayed on a
 logarithmic  scale,    overlayed  by    21~cm    continuum   contours
 \citep{km97}.  The contour  levels  are  0.1,0.2,0.4,0.8  and  1.6~Jy
 beam$^{-1}$.  b) 2MASS $K$-band image of the same region overlayed by
 contours of   $^{13}$CO  integrated  intensity \citep{cs98}.  Contour
 levels are  8,16,32 and 64 K km  s$^{-1}$. The square  boxes mark the
 observed regions, triangular symbols show the positions of associated
 UCHII  regions and the  star symbols represent  the IRAS sources that
 satisfy the \citet{wc89} criteria.}  \vfil}
\label{overlay}

\end{figure*}

\section{Observations and Data Reductions}

Near-Infrared (NIR) observations were made at the 3.8~m United Kingdom
Infrared Telescope  (UKIRT) with the  facility imager {\em UFTI}.  The
{\em UFTI} pixel scale measures 0.091$\arcsec$; the available field of
view  is   $\sim 90\arcsec$.  Photometric  observations  through $JHK$
broad band filters were secured for six  fields during three nights of
service observing (15 Sep  2000, 31 May 2001  and 12 Aug 2001).  Total
exposure times in the $J$,  $H$ and $K$ filters  were 180, 180 and 360
seconds respectively  on each of the  target fields. The seeing varied
between 0.32$\arcsec$ to  0.9$\arcsec$ with a  mean of 0.5$\arcsec$ in
the $K$-band  for  the photometric  data.   A nine  point (3$\times$3)
jittered observing sequence was executed to  obtain data that provided
final   mosaics     with      a    total     field     of   view    of
$\sim$115$\arcsec\times$115$\arcsec$. We note that the signal-to-noise
ratio at the edges  of these mosaics are  lower  than that within  the
central 90\arcsec area.  Standard data reduction  techniques involving
dark subtraction and median-sky-flat-fielding was applied.

We used tasks available   under {\sc iraf} for  photometric  analysis.
{\sc daofind} was used to identify  sources in each image. A threshold
equal  to four times  the mean  noise of each   image was found to  be
sufficient for  this purpose.   A psf model  was computed  by choosing
several  bright stars  well-spaced  out in  the image.  Photometry was
performed using {\sc daophot}. Aperture corrections were determined by
performing  multi-aperture    photometry  on  the     psf stars.   The
instrumental magnitudes were calibrated to   the absolute scale  using
observations of  UKIRT faint standards  (FS29,FS35 \& FS149) that were
obtained   at airmasses closest   to  the  target observations.    The
resulting photometric data are in the  natural system of the Mauna Kea
Consortium Filters \citep{st02}.  For  the purposes of  plotting these
data in Fig.4 and Fig.5 we have converted them  to the \citet{bb88}
(BB) system, since the main-sequence  references are in the BB system.
Since the direct transformation equations between the Mauna Kea system
and the BB sytem are not established as of now, we first converted the
Mauna  Kea system  to  CIT  system  and then to  the   BB system using
equations given by  \citet{haw01}.  We note  that this indirect method
can  induce high  errors  for  extremely red objects.   Representative
sub-images  of G48.9-0.3   and   G49.5-0.4, consisting of   stars  and
nebulosity,  were  chosen  for  determining the   completeness limits.
Limits  were established   by   manually  adding and  then   detecting
artificial stars of differing magnitudes.  By determining the fraction
of   stars  recovered in  each magnitude    bin,  we have deduced 90\%
completeness limits  of 18.2, 16.9  and 17.3  in the  $J$, $H$ and $K$
bands respectively.  Our  observations are absolutely complete (100\%)
to the level  of 16.2, 14.4 and  15.8  magnitudes in $J$, $H$  and $K$
respectively.  The relatively higher detection  level in $K$ is due to
the fact  that the exposure  times were twice that in   the $J$ \& $H$
bands. The  photometric accuracy of data  presented  here is generally
$<$0.2\,mag  although more than 30\% of  the  data points presented in
electronic    tables      have   magnitude      errors    less    than
0.05\,mag. Astrometry for  each of the  images was calibrated by using
the 2MASS as reference. The astrometric accuracy of the data presented
in this paper is better than 0.5\arcsec.

\begin{table}
 \centering   \begin{minipage}{100mm}    \caption{Summary  of observed
 fields} \label{photstat} \begin{tabular}{@{}llrrrrrr@{}}    \hline  &
 \multicolumn{2}{c}{Coords(J2000)}&PSF  &\multicolumn{4}{c}{Number  of
 Stars}\\ Name   & RA &  DEC &  \arcsec &J&H&K&JHK\\ \hline G48.9-0.3&
 19:22:14 & 14:03:09 &0.36  &662&1359&2071&593\\ G49.0-0.3& 19:22:26 &
 14:06:46 &0.50   &954&1659&1757&702\\ G49.2-0.3& 19:23:02  & 14:16:41
 &0.90  &469&882&905&293\\    G49.4-0.3&  19:23:05  &   14:28:07 &1.10
 &542&702&744&304\\    G49.5-0.4&  19:23:40    &  14:31:07       &0.32
 &651&977&1865&529\\
\hline
\end{tabular}
\end{minipage}
\end{table}

\begin{figure*}
\includegraphics[height=190mm]{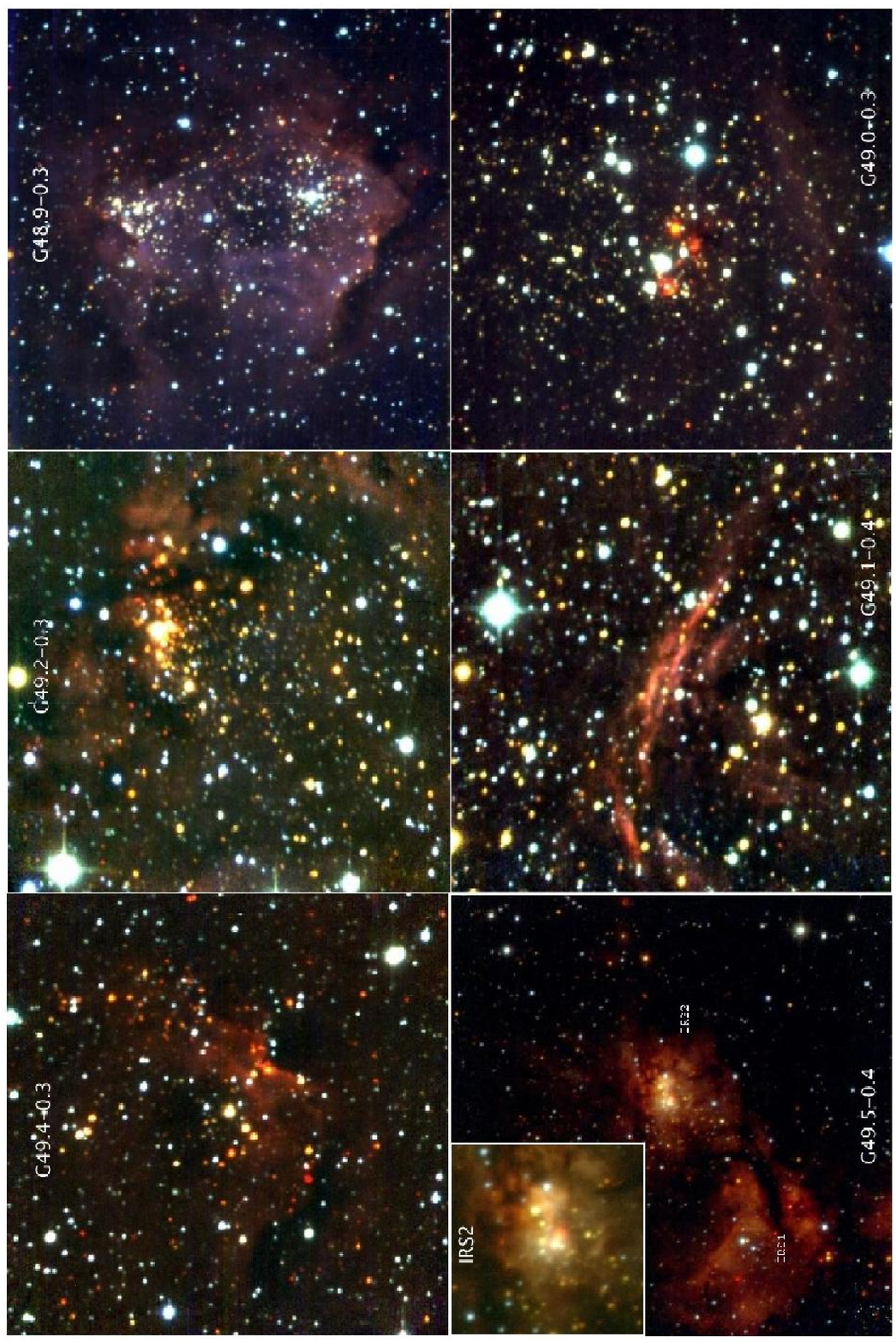}
\caption{JHK color composites of W51 Embedded clusters}
\end{figure*}

\section{Embedded clusters in W51 GMC}

We imaged  six fields along  the  W51GMC as  represented by  the bold,
square boxes in Fig.\,1.  The color plate in Fig.\,2 displays JHK band
three color composites  of each of the   observed fields. Regions  are
referred to using the nomenclature of the associated UCHII region. The
images are displayed on a linear  scale except for the field G49.5-0.4
which is shown with a logarithmic scale to  reveal the central region.
As can be seen from Fig.\,2, five of these six fields possess clusters
of stars immersed in $K$-band nebulosity.  The definition of a cluster
here implies: (1)  a significant increase in  the stellar density with
respect  to  an adjacent  background,  and  (2) association with radio
continuum sources.  The field around  G49.5-0.4 is associated with the
clusters known as W51 IRS1  and IRS2 \citep{gw94}.  The remaining four
clusters  are newly  discovered    from the  observations    presented
here. The star symbols in  Fig.\,1 mark the  positions of IRAS sources
that satisfy  the \citet{wc89} criteria for  UCHII  regions.  Three of
these IRAS  sources  are  associated   with  embedded  clusters;   the
remaining are unassociated with any NIR counterparts.  The geometrical
coincidence of the observed clusters  with the $^{13}$CO contours  and
the 21~cm contours indicate that these regions are ``embedded'' in the
gas and dust  of the W51~GMC.  Table.~1  summarizes  the positions and
statistics for the five fields,  derived from our JHK band photometric
analysis.   These numbers do not   reflect the actual number of  stars
that belong to   the clusters.  Our  photometric  observations provide
statistically   meaningful  data for only    four of the five observed
clusters; we therefore  present detailed analysis  only for these four
clusters.  The regions of G49.4-0.3 and  G49.1-0.4 are not included in
the detailed analysis presented below.

\subsection{Notes on individual clusters}

The four embedded clusters  discussed below are associated with bright
FIR   and radio continuum  emission.   They  are,  in their own right,
interesting objects for  detailed  high resolution studies at  various
wavelengths.

{\it G49.5-0.4}:  This  HII region  complex is  composed  of two major
components  known as  W51d and  W51e,  which  are associated  with two
embedded  clusters identified by  the prominent  infrared sources IRS2
and IRS1(see  Fig.\,1) The brightest  region, W51d, is well-studied at
centimeter, millimeter and infrared  wavelengths.  W51d/IRS\,2  is now
known  to be a young   embedded star cluster  that is  larger and more
luminous  than the  Orion-Trapezium cluster.   It  contains the  water
maser complex W51N \citep{forster89}, several OH and SiO maser sources
(\citet{caswell03};\citet{fish03}) and is thought to contain several O
type young stars.  In Fig.\,2, the  inset of the G49.5-0.4 panel shows
three-color   JHK-composite image  of  the    central region   of  the
W51d/IRS\,2   cluster which, at  a spatial  resolution of 0.32\arcsec,
reveals more  details than  any other  image of this  region  obtained
until now.  The region shows bright stars and nebulosity, the nebulous
cloud appearing  to  be  pinched off   to  form  two  lobes  stretched
east-west.   These are called  IRS2E and IRS2W  and  are shown to have
electron    densities  of  10$^5$-10$^6$  cm$^{-3}$\citep{oku01}.  Our
observations show that IRS2E is associated with a dense cluster, while
IRS2W is relatively devoid of stars. Further, the high-resolution view
suggests that IRS2E and  IRS2W are probably  two massive star  forming
clumps  bound  by  a  Roche-lobe  potential.    The central  region is
associated with 6-7 mid-infrared point-sources \citep{kra01, oka01}, 6
of   which appear as point sources   in our high-resolution NIR image.
One  of  the sources appears  as    a red nebulous  condensation  (the
brightest star  of the western blob  in the inset)  and corresponds to
the source KJD3   \citep{kra01} which is known  to  be associated with
deep silicate absorption features.  Most of these mid-infrared sources
are  thought to be O-type stars,  although there is  some debate about
whether they are O5-O6 or O9 type stars \citep{oka01}.

\begin{figure}
\vbox to120mm{\vfil
\includegraphics[width=85mm]{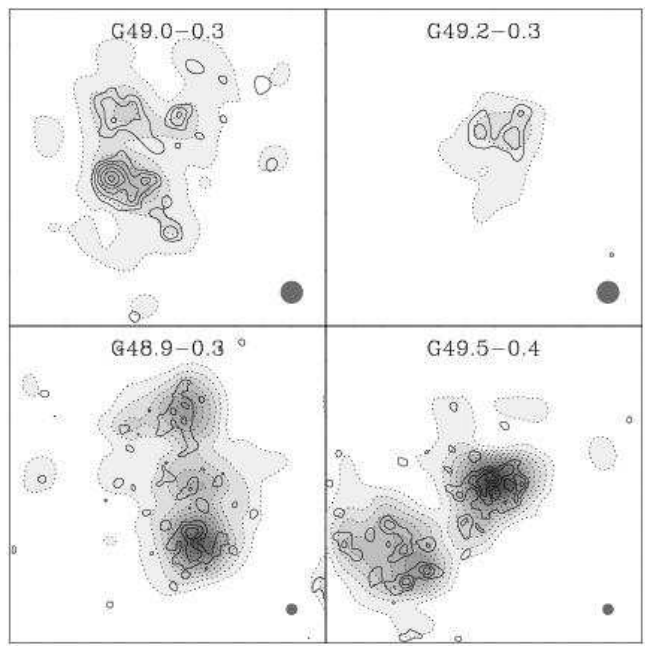}

\caption{Stellar surface density contours generated by Nyquist binning 
K-band star counts.  The dotted  curves with grey shades represent the
boundaries of  the clusters and  are obtained  by using  a binwidth of
7\arcsec.   The levels   of   the    dotted  contours  begins   at   4
stars/pixel(7$\arcsec$) and  increase in steps of  2 stars/pixel.  The
solid curves represent the hidden structure  of these clusters and are
revealed  by using smaller bin-widths  represented by the grey circles
in each panel.  }

\label{starden}
\vfil}
\end{figure}

{\it G48.9-0.3}:   This source  is  associated with  one of  the  most
spectacular clusters discovered in the W51GMC.  The top right panel in
Fig.\,2 display the three   color composite of this  newly  discovered
embedded  cluster.  This cluster  is located  towards one  edge of the
$\sim$100\,pc long 68\kms molecular cloud.  It appears to be placed on
the tip of the molecular cloud and the  observed HII region W51B. High
resolution  ISOCAM data for this  region  show strong FIR emission  at
6$\mu$m  with a morphology similar  to  the 8$\mu$m  MSX data shown in
Fig\,1.  This is  a twin cluster immersed  in  a cradle of  nebulosity
visible at  NIR wavelengths.  The southern  sub-cluster is a centrally
symmetric cluster  with  a   bright white/blue star  situated   at the
geometric center.  In contrast, the northern sub-cluster does not have
a  prominent     source,    although   it    does    appear  centrally
symmetric. Indeed, here there appears to be a small ring of stars that
is orientated in the plane of the sky.

{\it G49.2-0.3}: This is the strongest UCHII region  found in the W51B
region and corresponds approximately to  the location where the  shock
from the W51C supernova remnant is  thought to be interacting with the
W51  GMC \citep{km97}.  It is associated  with  a group of  bright red
stars located at  the boundary of a  cometary  shaped HII region.  The
K-band  (red) nebulosity has a similar  morphology and  is found to be
well  aligned   with  the    radio  free-free  emission    shown    in
Fig.\,1. Unfortunately,  our photometric data  for this region suffers
from relatively poor seeing (0.9$\arcsec$)  and a tracking error which
does not allow us to effectively resolve the  otherwise dense group of
faint sources that is enclosed by the cometary shaped HII region.
 
{\it G49.0-0.3}: The  bottom right  panel of  Fig.\,2 shows  the three
color composite of this field.  This region of the W51GMC is bright at
6-8$\mu$m and is relatively weak in terms of radio continuum emission,
probably indicating  a younger   cluster   and weaker HII    region in
comparison with  other brighter sources  in the W51B region.   A faint
``bow'' of K-band  nebulosity,  that is  similar in  morphology to the
associated radio continuum emission, surrounds  the cluster.  This  is
probably a photo-dissociation region.   Our three-color image shows  a
bright red star centrally-located in  the cluster.  Also found are two
knots of K-band (red) nebulosity in the central region that can not be
easily related to any particular star.

\subsection{Spatial distribution of stars}

It  is  evident  from   Figure.1b that  the  spatial  distribution  of
near-infrared sources is non-uniform, on  the global scale of the  GMC
and also on the local  scale of the individual  clusters.  In order to
study the nature  of this  distribution it  is  necessary to construct
star  count   surface   density   maps (SDMs).     Such  maps  provide
quantitative  estimates   of  stellar  surface   densities  and  their
variation  across the clusters.   They  also reveal details  about the
nature of  star formation and  the  hierarchical structure.  Different
methods of generating such maps have  been followed in various studies
of  nearby young star  clusters.  While \citet{lada95} use overlapping
rectilinear grids of boxes separated by  Nyquist criteria to count the
sources in the IC348 cluster, \citet{hh98} bin the source  counts into
10$\arcsec$  pixels and smooth  the   resulting map with  a  symmetric
gaussian  profile with   a 3-pixel  width to  produce  stellar surface
density maps of the Orion Nebula Cluster.

\begin{figure}
\vbox to130mm{\vfil
\includegraphics[width=95mm]{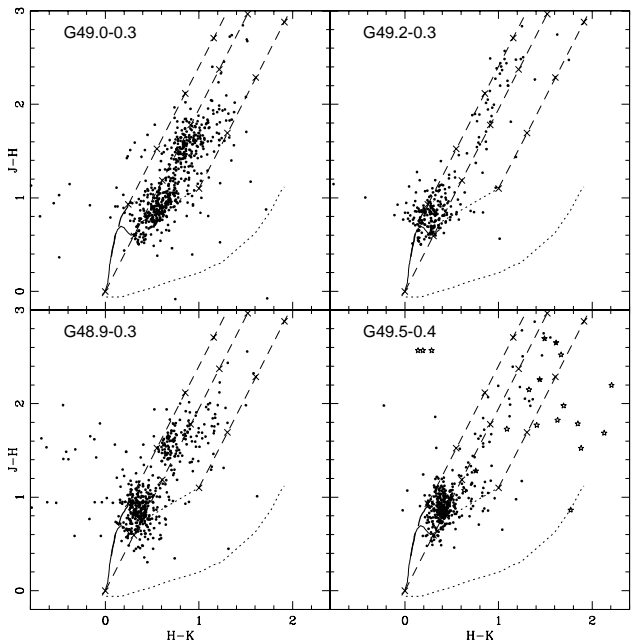}

\caption{Colour-colour diagrams for four clusters. The solid and broken  
heavy  curves  represents  the main-sequence dwarf    and giant stars,
respectively, and the dashed parallel  lines are the reddening vectors
that enclose    reddened  main-sequence objects. The  crosses   on the
reddening  vectors  are separated by   A$_v$=5\,mag\.  The dotted line
indicates  the locus of  T-Tauri stars  \citep{meyer97} and the dotted
curve represents the HAeBe locus  \citep{ladadams92}. The star symbols
in the  bottom  left  panel represent  the   sources in the   W51 IRS2
region. }

\label{2col}
\vfil}
\end{figure}

In  the  above cases the targets   of study were  nearby  star forming
regions within 500~pc distance where meaningful source counts could be
obtained over projected angular areas  of several arcminutes.  Massive
star forming regions like  W51 are generally  at much  large distances
and thus subtend relatively small angular sizes making it difficult to
study star count distribution. In these cases spatial resolution plays
an important role in the extraction of meaningful information from the
SDM.   The sizes of  individual features we may  expect  to see on the
SDMs depends on the Full-Width Half Maximum (FWHM) of the Point Spread
Function (PSF), the binning size adopted to  generate the SDMs and the
count  level   itself.  Together, these   determine  whether or  not a
meaningful SDM can  be made.  The binning size  is a  smoothing factor
that has to be  carefully chosen in  such a  case.  While a  small bin
size can include just a few sources and not  provide a meaningful SDM,
an overly  large bin size  can  hide the true   features in the  SDMs.
Given  the high quality of the  seeing (0.32$\arcsec$) in the $K$-band
images   of  two fields,   we experimented  with  different  bin sizes
corresponding to 6, 9, 12,  15, 18 and 22 times  the FWHM of the PSFs.
We found that while the majority of identifiable features did not vary
significantly between  6 and 12 times  FWHM, the features began to get
smeared for a bin size larger than 12 times FWHM. However, to identify
the  boundaries of  the overall   cluster and  estimate its   physical
characteristics,  a much larger bin  size, to smooth all the features,
had to be employed.

In Figure.~3 we show  SDMs for four  fields where the  detected source
counts were sufficient  to produce such a  map.  We  have adopted the
method of  \citet{lada95} who  use  a rectilinear grid  of overlapping
squares  separated by half the width  of individual squares to provide
Nyquist  sampling of the  sources.  We  counted  all sources from  the
$K$-band photometry  within  our completeness limits  discussed in the
previous section.  The cutoff  $K$ magnitude for  this choice was 17.8
(85\% detection limit); the  magnitude errors were less  than 0.3~mag.
The dotted  contours and    grey  shades in  Fig.~2  indicate   source
distribution as obtained  by a large smoothing  box (150 pixels $\sim$
14$\arcsec$).  These    contours  are used  to   identify  the cluster
boundaries  and estimate   the    average  surface density  of    each
cluster. The  thick contours overlayed  on  the grey shades  represent
SDMs generated by using    a smaller binsize (14 $\times$\,PSF)   that
shows smaller  groups within these  clusters.   All the four  clusters
reveal a  clumpy distribution of  the NIR  sources and the  clumps are
found to be  either elliptical or circular  in  shape.  The elliptical
clumps are particularly  noticeable   in the  maps of  G48.9-0.3   and
G49.5-0.4.  The projected dimensions of these clumps are approximately
0.1--0.2~pc.  Using  the SDM generated with  large bins (smooth maps),
we estimated the number of stars and the stellar surface densities for
each cluster listed in Table.\,2.

\begin{figure}
\vbox to110mm{\vfil
\includegraphics[width=95mm]{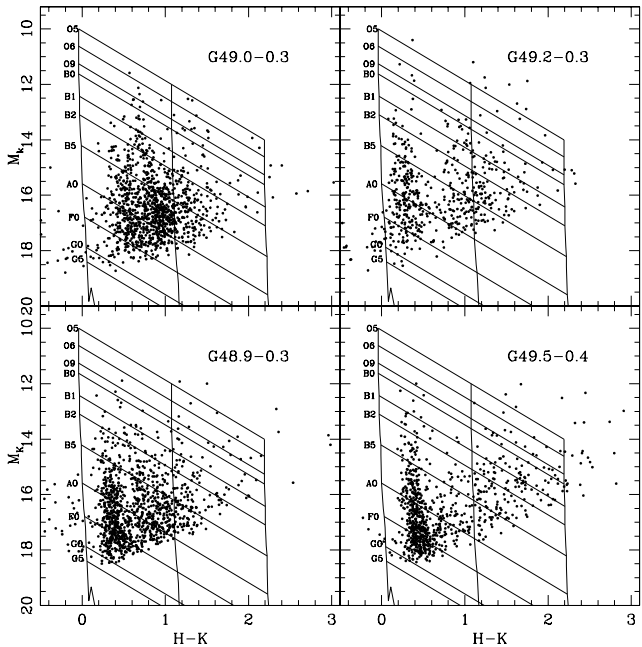}
\caption{Color-magnitude diagrams for four fields. The vertical solid 
lines (from left to right) represent  the main-sequence curve reddened
by 0, 20 and 40 magnitudes respectively. The horizontal parallel lines
represent reddening vectors.}
\label{absmagcol}
\vfil}
\end{figure}

\section{Color-Color(CC) and Color-Magnitude(CM) diagrams}

Table.1  lists the number of stars  detected in each of  the J,H and K
photometric bands, and simultaneously in  all three bands, for each of
the fields.  For the purposes of photometric analysis of the clusters,
we chose all  those stars with photometric  errors less then 0.2\,mag.
Tables.2-5 list the astrometry(RA, DEC) and photometry (magnitudes and
errors for  each of  the  J, H  and K  bands) of  the stars in sources
G48.9-0.3,G49.0-0.3,G49.2-0.3  and G49.5-0.4  respectively.  For   the
four clusters with SDMs  in Fig.\,3., the stars  detected in all three
photometric bands are  plotted on JHK  color-color diagrams in  Fig.4.
The solid  and broken heavy curves  represents the main-sequence dwarf
and  giant stars, respectively, and the  dashed parallel lines are the
reddening      vectors    that    enclose     reddened   main-sequence
objects.\citet{oku00} estimated the  slope of reddenning vector in the
direction of W51  to be $\sim$1.9.    Similar results are obtained  by
using   2MASS data.  We therefore  assumed  a  reddenning vector slope
E(J-H)/E(H-K)=1.9 (BB system) towards W51 region,  which relates to an
interstellar reddening law  with a value of  R=3.12 \citep{whittet80}.
The dotted line indicates  the locus of T-Tauri stars  \citep{meyer97}
and the dotted  curve  represents the HAeBe locus  \citep{ladadams92}.
Stars that lie outside  the  region of reddened main-sequence  objects
are  young stellar objects (YSOs)  with intrinsic  color excesses.  By
dereddening the stars   (on the  CC  diagrams)  that  fall within  the
reddening vectors encompassing the main  sequence stars and giants, we
found the visual extinction to each star.  We de-redenned the stars to
the  K6-M6 part of the  sequence  of stars.  The individual extinction
values range   from 0 to 14  magnitudes.   From a   histogram of these
values, we estimate the  average foreground extinction  to each of the
clusters.  Two of   the  clusters (G48.9-0.3  \&  G49.0-0.3)  show two
prominent peaks  of extinction on  the histogram corresponding  to the
two groups  of points that  can be identified on the  CC diagram.  The
estimated values of extinction are listed in Table.\,2.

It  can be noted  from Fig.\,4 that  there are significant differences
between the stellar population  of  the four clusters.  G49.0-0.3  and
G48.9-0.3 shows  two  distinct  groups of points    on the CC  diagram
separated by about 5-10  magnitudes of visual extinction.  In contrast
G49.2-0.3 and G49.5-0.4 do not show such  distinct groups. Nearly 90\%
of  the points  that   belong to   G49.0-0.3  lie  in the  region   of
intrinsically reddened objects,  thus signifying  a  high fraction  of
YSOs. Further, it  appears that G49.0-0.3 has two  groups of YSOs, one
group more reddened than the other.  We found that these two groups of
YSOs are  distributed homogeneously over the  cluster  area.  Thus the
more reddened group  is likely embedded  in the  molecular clump while
the least reddened group is  probably partially out  of or in front of
the  clump.  The bright  red star  in G49.0-0.3  (see Fig.\,2)  at the
center  of the cluster  shows colors similar  to a T-Tauri star and is
highly reddened  (A$_v \sim$35\,mag).   This could  be a massive  star
embedded in this cluster-forming core.

In G48.9-0.3 there  is a good  mixture of  both  main-sequence and YSO
populations.  However, unlike G49.0-0.3,  the two groups in  this case
are found to be distributed  differently.  The reddened group of stars
are all  found to occupy the  central  parts of the  two (northern and
southern) sub-clusters  (see   Fig.\,2),  while  stars  without   much
reddening  are  found to  be distributed  over the  entire  FOV of the
image.  Therefore the  average   extinction  value derived  from   the
reddened group  of  stars  is a   good  representation of the   visual
extinction to the molecular clump that hosts the two clusters.

The  G49.5-0.4  colour-colour   diagram   shows mostly   main-sequence
objects.  There is a small percentage  of YSOs with varying degrees of
reddening. The star symbols in Fig.\,4 mark the positions of the stars
associated with IRS2 (which is shown inset in the bottom-left panel in
Fig.\,2).  These stars are thought to  be O-type stars.  The placement
of these  sources in the upper-right  part of the  CC diagram supports
the conjencture that these are indeed O-type stars \citep{ladadams92}.
Finally, in  Fig.\,4 G49.2-0.3 displays very few  YSOs  and only a few
reddened main-sequence objects.  If  we consider the ratio of  sources
that fall in the main-sequence zone  and the YSO  zone as an indicator
of the cluster age, then the differences  in the CC diagrams discussed
above imply that  G49.0-0.3  is the   youngest cluster  in W51,  while
G48.9-0.3 is  the second youngest  cluster.   In comparison, G49.5-0.4
and G49.2-0.3 are relatively evolved objects.

\begin{table*}
 \centering
 \begin{minipage}{140mm}
  \caption{ Embedded Clusters in W51\,GMC }
  \label{embclusters}
  \begin{tabular}{@{}llrrrrrrr@{}}
  \hline
 & Number &\multicolumn{2}{c}{Stellar density (pc$^{-2}$)}& Radius(pc) &Age\footnote{age of associated HII region} & A$_v$ & Mass\footnote{Calculated for an age of 1Myr and A$_v\sim$7mag} \\
Name & of Stars & Average & Peak & ($\sqrt{A/\pi}$) & Myr & mag & \msun \\
  \hline
G48.9-0.3 & 658 & 175 & 475 & 3.43 & 1.4 & (5,10) & 10000 \\
G49.2-0.3 & 505 & 130 & 250 & 3.48 & 3.0 & (6,11) & 8000 \\
G49.4-0.3 & 148 & 123 & 175 & 1.95 & 2-3 & 5 & 2000 \\
G49.5-0.4(IRS1D)& 309 & 221 & 535 & 2.09 & 0.7 & 5 & 4400 \\
G49.5-0.4(IRS1E)& 333 & 179 & 340 & 2.39 & 0.7 & 5 & 4700 \\
\hline
\end{tabular}
\end{minipage}
\end{table*}

Fig.\,5 shows H-K vs K color-magnitude (CM) diagrams  for all the four
clusters. The vertical solid lines  (from left to right) represent the
main-sequence curve reddened by 0,  20 and 40 magnitudes respectively.
We  have assumed a  distance  of  6.5~kpc for  all  the  clusters, and
respective A$_v$ values  (listed in Table.1) to  each cluster in order
to  appropriately  reproduce the  main   sequence data on   this plot.
Similar to the  CC  diagrams, the CM  diagrams show  clear differences
between the four clusters.    The reddened MS  stars  are found to  be
spread parallel  to the  MS line.   Clusters G49.5-0.4,  G48.9-0.3 and
G49.2-0.3 clearly displays this concentration  of MS points, which  is
almost  completely absent in the  G49.0-0.3 cluster.  This agrees well
with the results  from the CC diagram  for G49.0-0.3, i.e.  that  YSOs
comprise about 90\% of the  total membership of  this cluster.  The CM
diagram  of  G48.9-0.3 shows  the YSOs  and   MS stars well separated.
Points that lie in the upper right corner of the CM diagram (above the
line corresponding to O5)  represent good candidates for massive YSOs.
The relative number of such  points are highest in G49.5-0.4, followed
by G49.2-0.3 and G48.9-0.3.  In  contrast, G49.0-0.3 has almost all of
its points concentrated  in the lower  right corner of the CM diagram,
corresponding to a population of essentially low mass stars.  However,
an estimation of the  true spectral type of each  star can not be made
using   the   CM  diagram  alone,   and   warrants  NIR  spectroscopic
observations of the individual sources.

\section{Discussion}

\subsection{Estimating Cluster Properties}

In Table.2  we summarise some of the  estimated properties of the four
embedded clusters in the  W51GMC.  These  estimates  are based on  the
photometric analysis presented in  previous sections and some standard
methods.  The clusters and their boundaries  were identified using the
SDMs shown in Fig.\,3.    The stars that   fall within  the  outermost
contours shown   in Fig.\,3  (dotted  contours  and grey shades)  were
counted to estimate the number of stars  in the cluster.  However, the
stellar  surface density (average and  peak  values) were estimated by
using the star counts within the contour equal to half the peak value.
It is   clear from  Fig.\,3  that   the clusters  are   not circularly
symmetric and indeed have irregular clumps.  The projected area of the
clusters were  used to define an  equivalent radius for the cluster as
R=$\sqrt{A/\pi}$, where A  is the estimated  area of each cluster from
Fig.\,3.  We have  estimated  a mass for each   cluster based  on  the
method described  by \citet{ll03}.    Following this method,   we have
assumed the K-luminosity function (KLF) models  of the Orion Trapezium
Cluster \citep{muench02} as   a  template and used    the evolutionary
tracks  of  \citet{dm94} in computing the   estimated masses.  We have
used  an age of 1Myr,   distance of 6.5Kpc   and an average foreground
extinction of A$_v$$\sim$7\,mag to compute the  masses.  The effect of
age on  the mass of the  cluster  is such that ages  $\le$1-2Myr yield
lower  masses compared to ages  $\ge$1-2Myr, with all other parameters
fixed.  However,  these assumptions are  fair, given the uncertainties
involved  in   our assumed parameters.    The estimated  mass  for the
G49.5-0.3  region (IRS1  and  IRS2)  together  is 9100\msun\  which is
similar   to  the mass estimated    by  \citet{oku00} for this  region
(8200\msun) who assumes a Salpeter mass function.  Our higher estimate
is as  expected because the spatial sampling  here is a factor  of 5-6
better compared  to \citet{oku00}, thus  revealing a larger  number of
cluster members.   We also note that  the  estimated masses  are quite
sensitive to the assumed extinction  values, particularly for  distant
sources such  as  these.   Therefore, the   estimated  masses can   be
uncertain by a few thousand solar masses.

The age of a   cluster is one   of the  most difficult parameters   to
estimate with   our present knowledge  because of   both observational
difficulties and  theoretical  uncertainties  (See \citet{ll03}  for a
discussion of   this problem).  In  the  case of massive  star forming
regions, some authors use the age of  the associated HII regions as an
indication of the age of the associated groups of stars. For example,
\citet{oku00}  give such   ages for  the four regions   falling in the
region of  W51A.  Extending this method to  the entire  W51GMC region,
one finds that the G48.9-0.3 is older  than G49.5-0.4(IRS2).  From the
studies presented here, if we  consider the ratio  of YSOs to MS stars
in each cluster  as an indicator of  its youth, and also consider  the
ages indicated by  the CM diagrams, it  turns out that the IRS2 region
is   older than the  G48.9-0.3  region.   This contradicts the  result
obtained by   comparing the  ages  of the  HII   region. It  has  been
suggested  that the UCHII  region phases of massive   stars last for a
time period much longer than 10$^5$yrs  if they are  born in dense and
hot regions  \citep{dePree}.  This is because  the pressure exerted by
the surrounding environment holds the   expansion of the UCHII  region
phase. It may even be possible that  massive star formation took place
at   a    much later   stage  than    the  low  mass   star  formation
(\citet{herbig62}; \citet{stahler85}).  Therefore, neither the ages of
the massive YSOs, nor the ages of the associated HII regions, are good
indicators of the age of the associated embedded clusters.

\subsection{Hierarchical and triggered star formation in W51GMC}

The studies presented here indicate the existence of at least two, and
possibly three, levels of hierarchy in the W51GMC.  From Fig.\,1 it is
evident that star formation has occurred in two major clumps, known as
W51A and W51B, within a $\sim$100pc-long molecular cloud. These clumps
need not be real and could indeed  be the effect  of disruption by the
supernova remnant W51C (see  \citet{km97} for a detailed  discussion).
The infrared image  of the W51A  region shows several  clumps that are
found as  associations of    stars  and nebulosity (see    Fig.\,1  of
\citet{hd02}). Next, as seen  in  Fig.\,2, there is clustering  within
each of the four clusters.  In particular, if  we consider the example
of G48.9-0.3, which   is a double   cluster, there  appears to   be an
additional  level   of hierarchy.  All   of  the clusters,  at various
levels, indicate hierarchical  clustering.   Hierarchical structure is
believed to be a direct consequence  of turbulent energy that operates
similarly   at   different     scales  (See,     e.g.,   reviews    by
\citet{elmegreen00}  \& \citet{mk04}).  Thus the observed hierarchical
clustering indicates the  presence of significant  turbulent energy in
W51GMC  which is not yet dissipated.   The observed star formation has
therefore taken place in the presence of strong turbulent energy.

If we sum-up  the mass of  all of the clusters  listed in Table.\,2 we
find a total    stellar mass of $\sim$30,000\msun.    Our observations
probably represents only $\sim$50\% of the total star formation in the
W51GMC, due to various reasons.  Firstly,  we have not sampled all the
smaller associations/clusters that can be seen  in a large scale 2MASS
K-band image. Next, our observations  have poor spatial resolution for
two   of the major  clusters; this  results  in an overly undersampled
cluster  membership for each  cluster. Also, the distributed component
of  individual   young  stars   across the cloud    can   contribute a
significant percentage to the stellar  mass which is neglected.  Thus,
the  total mass estimates from  these clusters alone are lower limits.
In view  of these uncertainties, it may  not be unreasonable to assume
that  the overall star formation  in the W51  GMC is associated with a
mass of $\sim$5-6$\times$10$^4$\msun.  In such  a case, the  estimated
mass of the W51GMC, $\sim$10$^6$\msun,   derived from observations  of
the molecular cloud itself (CS98), implies a star formation efficiency
(ratio of stellar mass to cloud mass) of $\sim$5-6\%. CS98 showed that
four of  the  radio  continuum sources, namely  G48.9-0.3,  G49.0-0.3,
G49.1-0.4  and G49.2-0.3,  are spatially  and kinematically associated
with the 68\kms cloud  alone.   If we believe   that all the  observed
clusters  in W51 GMC do indeed  belong to the   68\kms cloud, then the
mass estimate  of   $\sim$2$\times$10$^5$\msun\ for the   68\kms cloud
(CS98)  ,  together with    the mass estimates  for the    clusters in
Table.\,2, indicates star  formation efficiencies $\sim$10-15\%.  This
value  is  comparable to  the star   formation   efficiencies in  some
relatively   nearby massive star    forming clusters, such  as NGC6334
(25\%)\citep{tapia96} and W3 IRS5 (6-18\%) \citep{Megeath96}.

Low mass star formation in the solar neighbourhood is  known to have a
low efficiency,  of  the order  of 1-3\%.    Triggering mechanisms are
thought to result in the formation of clusters;  they are also thought
to  be  associated with  higher star  formation  efficiencies.  In the
W51GMC,    the  embedded    clusters     are   distributed  inside   a
$\sim$100pc-long cloud.   The    clusters also likely  have  ages   of
0.5-3\,Myrs  because that is   the commonly observed ages of  observed
embedded clusters \citep{ll03}.  This suggests near-simultaneous
star formation across the entire cloud, an event which could have been
spontaneous   and/or  triggered  by   an  external   mechanism.  If  a
triggering mechanism  was involved then it  would have to  have been a
large scale shock wave to influence the 100\,pc-long region.  Internal
trigger mechanisms are very unlikely in this  case, since the crossing
time   for the  cloud, for   an average  velocity of  $\sim$10\kms, is
$\sim$10\,Myr, which is larger than the  typical life time of 3-5\,Myr
for the embedded clusters.  When we consider these  facts and note the
location of the W51GMC at the  tangent point of the Sagittarius spiral
arm \citep{burton70}, it seems likely that the observed star formation
in the W51GMC  is triggered by  galactic spiral density  waves.  Giant
clouds like  W51 are indeed  thought to undergo gravitational collapse
under  the   influence of  spiral   density waves \citep{elmegreen94}.
Therefore, W51GMC  may be one of  the best examples   in our galaxy of
triggered star formation by spiral waves.

\vskip -5cm

\section{Summary and Conclusions}

We    present  sub-arcsecond  resolution,   near-infrared  J,H,K  band
photometric observations of six fields  in the W51 GMC (which includes
the W51A and  W51B regions). We  construct a multi-wavelength view  of
the W51 GMC, combining   CO  and $\lambda$21cm observations from   the
literature with MSX and 2MASS data. All of these observations together
suggest a unified view for W51, the  HII regions being associated with
a $\sim$100\,pc long molecular   cloud named the ``68\kms''  cloud  by
\citet{cs98}. The results  of  our near-infrared observations  can  be
summarised as follows.

1)From  our observations of six   regions, four new, embedded clusters
are discovered  that are associated with  the UCHII regions G48.9-0.3,
G49.0-0.3,  G49.2-0.3  and  G49.4-0.3.     Among these,  the   cluster
associated with G48.9-0.3 is found to be a twin  cluster enclosed in a
nest  of  near-infrared nebulosity.   Our observations  also provide a
sub-half arcsecond view   of the W51  IRS2 (G49.5-0.4)  region that is
known to be forming several O-type  stars.  This high resolution image
provides a view  which suggests that the  W51 IRS2E and  IRS2W are two
massive star forming clumps that are likely  connected by a Roche-lobe
potential.

2) Using K-band star counts, we construct stellar surface density maps
for  four  clusters, (G48.9-0.3, G49.0-0.3,  G49.2-0.3 and G49.5-0.4.)
These maps  unveil the underlying hierarchical  structure and allow us
to estimate some physical parameters for the clusters. Color-color and
color-magnitude diagrams for each   of the clusters indicate  distinct
differences in    their  embedded stellar   population.   The clusters
associated with  G48.9-0.3 and G49.0-0.3  are found to contain  a high
percentage of YSOs, suggesting their relative  youth in comparision to
the other regions studies.

3) Assuming an Orion  Trapezium IMF  and theoretical pre-main sequence
tracks, we   estimate  the masses of  the  individual  clusters.   The
combined mass of  all the clusters  amounts to a  net stellar  mass of
$\sim$10$^4$\msun\ in the W51 GMC. This estimate, when combined with
estimates  of the  molecular  cloud mass from  \citet{cs98}, suggests a
star formation efficiency of $\sim$10\%. 

Finally,  the combined view of the  W51 GMC, from our new observations
and from  the literature, suggests that star  formation in the W51 GMC
is  probably triggered by a  galactic  spiral density wave.  Thus, W51
may represent a relatively unique example of massive star formation in
our galaxy.

\section{Acknowledgments}
The  United Kingdom Infrared    Telescope  is operated  by  the  Joint
Astronomy Centre on behalf of the U.K.  Particle Physics and Astronomy
Research Council.  The UKIRT data reported  here were obtained as part
of its Service  Programme. We thank  John Carpenter  for providing the
$^{13}$CO data and Bon-Chul  Koo  for providing the  HI 21cm  data for
overlays presented in Fig.\,1. We also thank  the referee Rob Jeffries
for   useful   suggestions.  This  work  was    supported  by a  grant
POCTI/1999/FIS/34549 approved by  FCT and POCTI,  with funds  from the
European community programme FEDER.  This  research  made use of  data
products  from the Midcourse Space Experiment  and from the Two Micron
All Sky Survey. These  data products are provided  by the services  of
the Infrared  Science Archive operated  by the Infrared Processing and
Analysis Center/California  Institute  of  Technology, funded  by  the
National Aeronautics and Space Administration and the National Science
Foundation.

\label{lastpage}


\begin{thebibliography}{99}

\bibitem[\protect\citeauthoryear{Bessell \& Brett}{1988}]{bb88}
Bessell, M. S., Brett, J. M., 1988, PASP, 100, 1134

\bibitem[\protect\citeauthoryear{Burton}{1970}]{burton70}
Burton, W. B., 1970, A\&AS, 2, 291

\bibitem[\protect\citeauthoryear{Carpenter \& Sanders}{1998}]{cs98} 
Carpenter, J. M., Sanders, D. B., 1998, AJ, 116, 1856 (CS98)

\bibitem[\protect\citeauthoryear{Caswell}{2003}]{caswell03}
Caswell, J. L., 2001, MNRAS, 341, 551

\bibitem[\protect\citeauthoryear{D\'Antona \& Mazzitelli}{1994}]{dm94} 
D\'Antona, F., Mazzitelli, I., 1994, ApJS, 90, 467

\bibitem[\protect\citeauthoryear{de Pree}{1995}]{dePree}
de Pree, C. G., Rodriguez, L. F., Goss, W. M., 1995, RMxAA, 31, 39

\bibitem[\protect\citeauthoryear{Elmegreen}{1994}]{elmegreen94}
Elmegreen, B. G., 1994, ApJ, 433, 39

\bibitem[\protect\citeauthoryear{Elmegreen et al.}{2000}]{elmegreen00}
Elmegreen,  B. G., Efremov, Y., Pudritz,  R., Zinnecker,  H., 2000, in
Mannings, V., Boss, A. P., Russell, S. S.,  eds, Protostars \& Planets
IV, University of Arizona Press, p. 179

\bibitem[\protect\citeauthoryear{Fish et al.}{2003}]{fish03}
Fish, V. L., Reid, M. J., Argon, A. L., Menten,K. M., 2003, ApJ, 596, 328

\bibitem[\protect\citeauthoryear{Forster \& Caswell}{1989}]{forster89}
Forster, J.R., Caswell,J.L., 1989, A\&A, 213, 339

\bibitem[\protect\citeauthoryear{Goldader \& Wynn-Williams}{1994}]{gw94} 
Goldader, J.D., Wynn-Williams, C.G., 1994, ApJ, 433, 164

\bibitem[\protect\citeauthoryear{Hawarden et al.}{2001}]{haw01}
Hawarden, T. G.,  Leggett, S. K., Letawsky,  M. B., Ballantyne, D. R.,
Casali, M. M., 2001, MNRAS, 325, 563

\bibitem[\protect\citeauthoryear{Herbig}{1962}]{herbig62}
Herbig, G. H., 1962, ApJ, 135, 736

\bibitem[\protect\citeauthoryear{Hillenbrand \& Hartmann}{1998}]{hh98} 
Hillenbrand, L. A., Hartmann, L. W., 1998, ApJ, 492, 540

\bibitem[\protect\citeauthoryear{Hodapp \& Davis}{2002}]{hd02} 
Hodapp, K. W., Davis, C. J., 2002, ApJ, 575, 291

\bibitem[\protect\citeauthoryear{Koo}{1999}]{koo99} Koo, B., 1999, 
ApJ, 518, 760

\bibitem[\protect\citeauthoryear{Koo \& Moon}{1997}]{km97} 
Koo, B. C., Moon D. S., 1997, ApJ, 475, 194

\bibitem[\protect\citeauthoryear{Kraemer et al.}{2001}]{kra01} 
Kraemer,  K. E., Jackson, J.  M., Deutsch, L.   K.,  Kassis, M., Hora,
J. L., Fazio, G. G., Hoffmann, W. F., Dayal, A., 2001, ApJ, 561, 282

\bibitem[\protect\citeauthoryear{Lada \& Adams}{1992}]{ladadams92}
Lada, C. J., Adams, F. C., 1992, ApJ, 393, 278

\bibitem[\protect\citeauthoryear{Lada \& Lada}{2003}]{ll03} 
Lada, C. J., Lada, E. A., 2003, ARAA, 41, 57

\bibitem[\protect\citeauthoryear{Lada \& Lada}{1995}]{lada95} 
Lada, E. A., Lada, C. J., 1995, AJ, 109, 1682


\bibitem[\protect\citeauthoryear{Mac Low \& Klessen}{2004}]{mk04} 
Mac Low, M., Klessen, R. S., 2004, RvMP, 76, 125

\bibitem[\protect\citeauthoryear{Meyer et al.}{1997}]{meyer97} 
Meyer, M. R., Calvet, N., Hillenbrand, L. A., 1995, AJ, 109, 1682

\bibitem[\protect\citeauthoryear{Megeath et al.}{1996}]{Megeath96}
Megeath, S. T., Herter,  T., Beichman, C.,  Gautier, N.,  Hester, J. J.,
Rayner, J., Shupe, D., 1996, A\&A, 307, 775

\bibitem[\protect\citeauthoryear{Muench et al.}{2002}]{muench02}
Muench, A. A., Lada, E. A., Lada, C. J:, Alves, J. F., 2002, ApJ, 573, 366

\bibitem[\protect\citeauthoryear{Mooney et al.}{1995}]{msm95} Mooney, T., 
Sievers, A., Mezger, P. G., Solomon, P. M., Kreysa, E., Haslam, C. G. T., \& 
Lemke, R., 1995, A\&A, 299, 869

\bibitem[\protect\citeauthoryear{Okumura et al.}{2000}]{oku00} 
Okumura,  S.,  Mori, A., Nishihara,  E.,  Watanabe, E., Yamashita, T.,
2000, ApJ, 543, 799

\bibitem[\protect\citeauthoryear{Okumura et al.}{2001}]{oku01} 
Okumura,  S.,  Mori, A.,  Watanabe, E.,  Nishihara, E., Yamashita, T.,
2001, AJ, 121, 2089

\bibitem[\protect\citeauthoryear{Okamoto et al.}{2001}]{oka01} 
Okamoto, Y. K., Hirokazu, K., Takuya, Y.,  Takashi, M., Takashi, O.,
2001, ApJ, 553, 254

\bibitem[\protect\citeauthoryear{Simons \& Tokunaga}{2002}]{st02}
Simons, D. A., Tokunaga, A. 2002, PASP, 114, 169

\bibitem[\protect\citeauthoryear{Stahler}{1985}]{stahler85}
Stahler, S. W., 1985, ApJ, 293, 207

\bibitem[\protect\citeauthoryear{Tapia et al.}{1996}]{tapia96}
Tapia, M., Persi, P., Roth, M., 1996, A\&A, 316, 102 

\bibitem[\protect\citeauthoryear{Whittet \& van Breda}{1980}]{whittet80}
Whittet, D. C. B., van Breda, I. G., 1980, MNRAS, 192, 467

\bibitem[\protect\citeauthoryear{Wood \& Churchwell}{1989}]{wc89}
Wood, D. O. S., Churchwell, E., 1989, ApJ, 340, 265

\end{thebibliography}
\end{document}